\documentclass[pra,showpacs,twocolumn,eqsecnum]{revtex4}
\usepackage{dcolumn}% Align table columns on decimal point
\usepackage{bm}% bold math
\usepackage{amssymb}
\usepackage{amsmath}
\usepackage{amsfonts}
\usepackage[]{graphicx}
\begin{document}
\bibliographystyle{prsty}
\title{Spatio-temporal second-order  quantum correlations of surface plasmon polaritons}
\author{ M. Berthel, S. Huant,  A. Drezet $^{1}$}
\address{(1)Institut N\'{e}el, CNRS and Universit\'{e} Joseph Fourier,
BP166, 38042 Grenoble Cedex, France}
\begin{abstract}
We present an experimental methodology to observe spatio-temporal second-order quantum coherence of surface plasmon polaritons which are emitted by nitrogen vacancy color centers attached at the apex of an optical tip. The approach relies on leakage radiation microscopy in the Fourier space and we use this approach to test wave-particle duality for surface plasmon polaritons. 
\end{abstract}
\maketitle
\begin{figure}[htbp]
\centering
\fbox{\includegraphics[width=\linewidth]{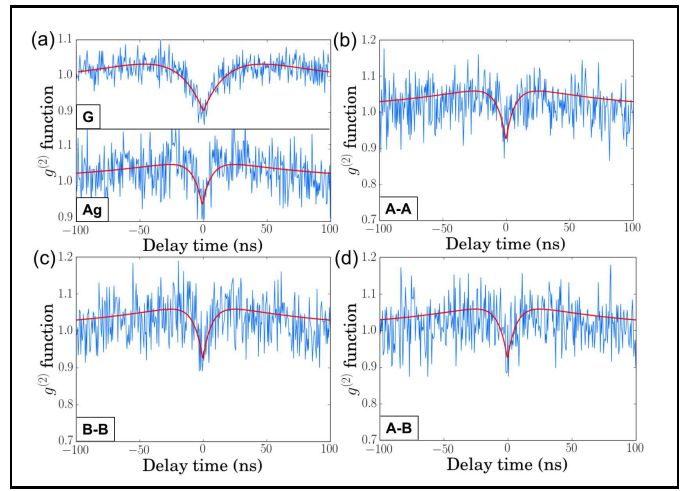}}
\caption{Second-order correlation functions in four different configurations. (a) G: tip over glass, compared to Ag: tip over the Ag film. (b-c) Auto-correlation in the Fourier plane for respectively the  A-A and B-B configuration. (d) both fibers collect SPPs emitted in different regions A-B of the Fourier space. Red curves correspond to a three-level system model. }
\label{G2}
\end{figure}

\indent Quantum plasmonics, i.e., the study of surface plasmon polariton (SPP)\cite{Barnes} interactions in the quantum regime~\cite{Maier}, has generated a growing interest in recent years due to the huge potentialities it offers to quantum information processing and integrated nano-photonics~\cite{Altewisher,Fasel,Martino1,Martino2}. Within this context, one of the most urgent concern is the optimization of the coupling with single SPP quanta of individual fluorescent emitters, such as quantum dots~\cite{Akimov,Fedutik,Wei} or nitrogen vacancy (NV) color centers in nanodiamonds (NDs)~\cite{Kolesov,Schie,Schell}, confined in planar nanostructures. Among the various methods for studying this regime, leakage radiation microscopy (LRM)~\cite{DrezetMatResB2008} coupled to near-field scanning optical microscopy (NSOM) is particularly promising due to recent progress in the fabrication of active tips with single quantum emitters glued at their apex\cite{Yannick,Cuche2009,revue}. Working with such NV-based NSOM tip it has been possible to launch single SPPs whose propagation along a thin metal film was mapped using LRM together with confocal microscopy and NSOM techniques~\cite{CucheNL2010,MolletPRB2012}. Subsequently it has been possible to prove that the photon-plasmon emission statistics, recorded using a time-resolved Hanbury Brown and Twiss (HBT) second-order interferometer coupled to the NSOM/LRM optical set-up, was fully conserved during the conversion back and forth of photons to SPPs and SPPs to photons~\cite{MolletPRB2012}.\\
\indent In this work we go a further step forward and demonstrate how to use NV-based NSOM coupled to LRM to measure not only the temporal but also the spatio-temporal second-order quantum coherence $g^{(2)}(\tau,\mathbf{k}_A,\mathbf{k}_B)$  of leaky SPPs emitted by few NVs in two different directions $\mathbf{k}_A$ and $\mathbf{k}_B$ in the LRM Fourier space and at two different times separated by $\tau$. Here, we experimentally demonstrate a proof of principle and show how the recipe could be used to realize quantum optics correlation experiments with combined LRM and NSOM.\\
\indent The setup used in this work is a home-built transmission NSOM working in ambient conditions and coupled to an inverted confocal microscope  and to a LRM setup. A chemically etched sharp tip obtained from a single mode optical fiber is glued onto a quartz tuning fork. The surface is probed by using a piezoelectric system and an electronic feedback loop. A $cw$ 532 nm diode laser is injected into the fiber to optically excite a tiny subwavelength volume surrounding the tip apex. The sample is imaged in the far-field through an oil immersion objective with a numerical aperture $NA$=1.4. The sample considered in this study is a 170 $\mu$m thick glass substrate, which is half covered with a 50 nm thick Ag film. On the other half part of the sample a drop of a solution containing 80 nm diameter NDs hosting few NVs is deposited. After evaporation of the solvent we can optically characterize with a HBT interferometer the intensity/time second-order correlation $g^{(2)}(\tau)$ of the emission of a single nanocrystal. The later is preselected using confocal optical microscopy (the complete procedure is reviewed in ref.~\cite{revue}). We remind that $g^{(2)}(\tau)$ is defined as 
\begin{equation}
% \left
  g^{(2)}(\tau)=\frac{P(t+\tau,t)}{P(t)^2}=\frac{P(t+\tau \vert t)}{P(t)},
  \label{g2prob}
% \right
\end{equation} 
where $P(t+\tau \vert t)$ is the conditional probability (per unit time) to detect a photon at time $t+\tau$ knowing that a photon was detected at time $t$, and $P(t)$ is the normalization probability (per unit time) to detect a photon at time $t$.  In particular for $N$ independent quantum emitters acting as single photon sources, $g^{(2)}(0)$ is defined as~\cite{beveratos,CucheNL2010} : 
\begin{equation}
% \left
  g^{(2)}(0)=1-\frac{1}{N}.
  \label{g2}
% \right
\end{equation} 
This equation is important in our context as we mostly work with NDs hosting several NV centers in order to have enough detection signal. However, we can still have a quantum view of the phenomenon because the system cannot emit more that $N$ photons at a time. Furthermore, the fact that $g^{(2)}(0)<1$ is a specific quantum signature which has no counterpart in the classical electromagnetic world. In the following we will work with $N=10$ as shown later.\\  
\indent In the next step we approach the NSOM tip from the preselected nanodiamond in order to pick it up from the surface using a protocol presented in refs.~\cite{Cuche2009,CucheNL2010,MolletPRB2012}. Briefly, a positively-charged polymer is deposited on the tip apex to make the ND grafting easier. When the tip is scanned over the selected ND, the tip-surface distance is reduced to 10-20 nm to force the contact between the tip and the crystal. The grafting can be directly checked by measuring the $g^{(2)}(\tau)$ function of the fluorescence signal emitted by the active tip. 
After the microscope objective, the optical signal is cleared from the remaining transmitted  excitation by using a dichroic mirror and a long-pass dielectric filter with $\lambda_{cut-on}=580nm$, and is subsequently sent \begin{figure}[htbp]
\centering
\fbox{\includegraphics[width=\linewidth]{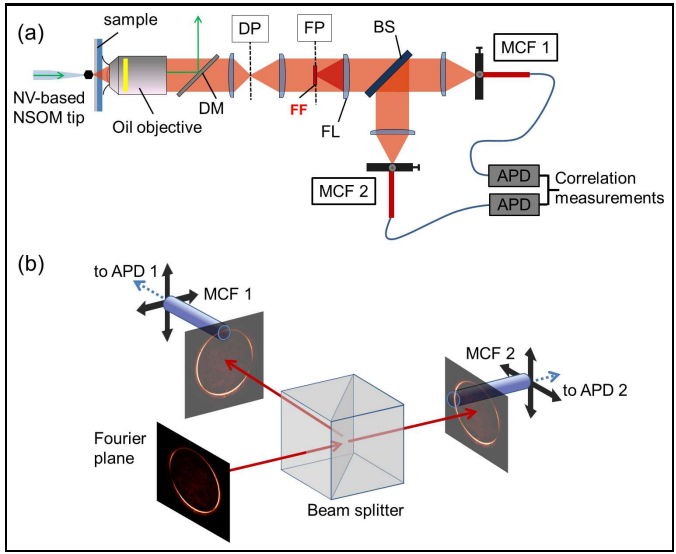}}
\caption{Sketch of the spatial HBT correlator. (a) Complete view of the setup. A dichroic mirror (DM) filters the excitation light (green arrows). Either the direct plane (DP) or the Fourier plane (FP) can be observed by removing or adding the Fourier lens (FL), respectively. A Fourier filter (FF) is added is the FP to block spatially the allowed, i.e. non plasmonic, light. Finally, a beam splitter (BS) directs
the light into two motorized collection fibers (MCF1 and 2), and the
photons are counted using APDs. (b) Schematic view of the MCFs in the case of a plasmonic circle in the Fourier plane. }
\label{HBT}
\end{figure}to either a charge-coupled device (CCD) camera, a spectrometer or an HBT correlator for analysis. In this work we additionally implement a specific HBT correlator adapted to the recording of $g^{(2)}(\tau,\mathbf{k}_A,\mathbf{k}_B)$ by mixing HBT and LRM methods. We remind that LRM is mainly a far field microscopy method relying on the fact that through a sufficiently thin metal film SPPs propagating at the air/metal interface can leak in the substrate at a large angle $\Theta_{\textrm{LRM}}$, larger than the critical angle for glass $\Theta_c$~\cite{DrezetMatResB2008}. Using an oil immersion objective it is thus possible to map the SPP propagation in either the direct or Fourier space. In ref.~\cite{CucheNL2010} LRM was used to record SPP launched from a NV based NSOM tip and the $g^{(2)}(\tau)$ of quantized SPPs was measured by using a collection optical fiber located in the Fourier plane and connected to a standard HBT system to show the preservation of the second-order coherence in the time domain~\cite{MolletPRB2012}. Here, in order to measure spatio-temporal correlations we precisely locate in the Fourier plane of the LRM two collection optical fibers  mounted on motorized stages that are directly plugged to avalanche photodiodes (APDs). Hence, the collection fibers can scan up to 1 $mm^2$ on the observed plane. Furthermore, we can alternatively image the Fourier or the direct space thanks to a removable lens inserted in the optical path.  We can therefore perform spatial correlation measurement between two different points in either the  direct plane or in the Fourier plane. In the present work we will focus on correlation in the Fourier space, i.e, on  $g^{(2)}(\tau,\mathbf{k}_A,\mathbf{k}_B)$.  Fig.~\ref{HBT} is a sketch of this spatial HBT correlator in the case of Fourier plane imaging.\\ 
\indent We now present the experimental results obtained with a ND hosting $N=10$ NV centers . It is challenging to go far below this number as the measurement involves a detection of very low SPP signals. A NV center itself is bright (count rate 10-20 $kHz$), when measured through the glass substrate only. But spatial correlation measurement of SPP implies some losses on the photon path: First, only a fraction of the NV fluorescence is coupled to SPPs, and then most of it is absorbed by the silver film during propagation. Finally, only a small part of the leaking SPPs is detected using the movable collection fibers. With a ND containing 10 NV centers, this results in a $g^{(2)}$ function measurement with count rates of about 5-10 $kHz$ per APD.\\
\begin{figure}[htbp]
\centering
\fbox{\includegraphics[width=\linewidth]{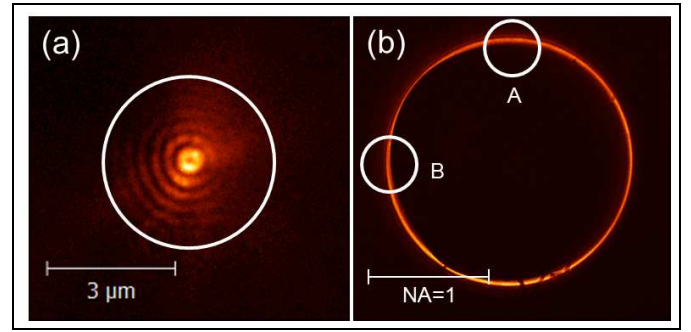}}
\caption{CCD images of the ND-based tip facing the Ag film, at a distance around 40 nm. The two images have been obtained after a 1 min exposure time. (a) Direct plane. A Fourier mask obstructs direct light so that we only see leaking plasmons. The white circle corresponds to the collection fiber diameter. (b) Back focal plane. The emitted SPPs give rise to a circle at $NA_{\textrm{SPP}}=1.04$, that can be isolated from low $NA$ light with the mask mentionned in (a).  A and B white circles show the location and relative size of the collection fibers.}
\label{CCD}
\end{figure}
\indent After grafting the ND, we approach the tip to 40 nm above the Ag film. This near-field configuration permits the NV fluorescence to couple to SPPs. Fig.~\ref{CCD} shows the CCD imaging of the direct (a) and back-focal (b), i.e. Fourier, planes. A mask filters direct light below the critical angle $\Theta_c$, defined as $NA_c=n_g\sin{\Theta_c}=1$ (with $n_g\simeq1.5$ the glass optical index) so that Fig.~\ref{CCD}(a) only shows the leaking SPPs cleared  from other contributions. In the direct plane, according to theory~\cite{PRL}, we see a decaying interference pattern from the center that corresponds to the SPP point-spread function of the LRM centered on the tip position.  We observe a "doughnut" shape at the center, which is typical of a vertical dipole configuration~\cite{PRL,BharadwajPRL2011}. The ND containing 10 NV centers, we can assume that it corresponds to a distribution of arbitrary oriented dipoles, i.e, to a mixture of vertical and horizontal components. Furthermore, it has been shown ~\cite{PRL,Martin1} that the vertical dipole coupling efficiency to SPPs is much higher than for horizontal dipoles by a factor up to $\eta=|k/k_z|^2\simeq 15$, where $k$ and $k_z$ are the in-plane and out of plane components of the SPP wavevector on the air side, respectively. The experiment confirms qualitatively this finding. The white circle in Fig.~\ref{CCD}(a) is a representation of the collection fiber area, which physically has a diameter of 200 $\mu$m but which, due to the microscope magnification, is here reduced to 4 $\mu$m. In this configuration we can record the $g^{(2)}(\tau,\mathbf{x},\mathbf{x})$ function where $\mathbf{x}$ is the tip position. In the Fourier plane, i.e., Fig.~\ref{CCD}(b), the leaking SPP profile draws a ring of radius $NA_{\textrm{SPP}}=n_g\sin{\Theta_{\textrm{LRM}}}\equiv n_{\textrm{SPP}}=1.04$ which is specific of the SPP effective index $n_{\textrm{SPP}}$ at the considered optical wavelength (i.e., $\lambda\simeq$ 650-750 nm)~\cite{DrezetMatResB2008}. The white circles labeled by A and B indicate the positions at which we performed the spatial correlation measurements in the Fourier plane. They permit to illustrate the relative size of the collection fiber core compared to the plasmonic circle. Each fiber collects barely 7\% of the total SPP signal as estimated from the ratio between the collection area diameter to the SPP ring perimeter (which takes into account the magnification of the optical setup).\\
\indent In the final step of the experiment, we record the spatio-temporal second-order quantum coherence for various configurations. The experimental results are shown in Fig.~\ref{G2}. We first measure the usual $g^{(2)}(\tau,\mathbf{x},\mathbf{x})=g^{(2)}(\tau)$ in the direct plane (see Fig.~\ref{G2}). Since the SPP point-spread function, which mimics approximately the actual SPP field on the air/metal interface, decays exponentially with the in plane distance $\rho$ away from the tip, see Fig.~\ref{CCD}(a), we here limit the analysis to auto-correlation measurements when the ND is above the glass substrate, as a reference measurement, and when the ND is facing the silver film, with the Fourier filter in place. For comparison the experiments on the silver film are done with and without the Fourier filter, showing significant difference. In the back-focal plane, we perform spatial measurements of auto-correlation (both collection fibers are in point A or B as shown in Fig.~\ref{G2}(b,c)) and cross-correlation (one fiber is on A and the other on B, as shown in Fig.~\ref{G2}(d) ) corresponding to $g^{(2)}(\tau,\mathbf{k}_A,\mathbf{k}_A)$, $g^{(2)}(\tau,\mathbf{k}_B,\mathbf{k}_B)$ and $g^{(2)}(\tau,\mathbf{k}_A,\mathbf{k}_B)$, respectively.  The first point to highlight is that all curves show an antibunching dip  at $\tau=0$, which is a clear signature of the quantum nature of light. The $g^{(2)}$-curves on the metal film are also more noisy than on glass due to the lower signal that reaches the APDs. 

Before going to the quantitative analysis, we already clearly see in Fig.~\ref{G2} that the antibunching dip is much narrower when the tip is facing silver than when it is facing glass. This can be due either to an increase of the pump power or a rise of the intrinsic spontaneous emission rate. \\
\indent We use a fit routine to analyze these curves. The detailed method is described in ref.~\cite{Martin2}. In short, we use a three-level system to model the NV center dynamics, with a ground level (1), an excited level (2) and a intermediate metastable level (3). Four channels are taken into account : the excitation channel of rate $k_{12}$, the spontaneous emission channel of rate $k_{21}$ and the two non-radiative channels with coefficients $k_{23}$ and $k_{31}$. The Einstein population equations lead to the following $g^{(2)}$ formula : 
\begin{equation}
% \left
   g^{(2)}(\tau)=1-(\beta e^{-\gamma_1 \tau}+(\beta-1)e^{-\gamma_2 \tau})\frac{\rho^{2}}{N},
  \label{g2}
% \right
\end{equation}
where $\rho$ is the signal-to-noise ratio, and $\gamma_1$, $\gamma_2$, $\beta$ are defined by~\cite{Martin2}:
\begin{gather}
\gamma_1\simeq k_{12}+k_{21},\\
\gamma_2\simeq k_{31}+\dfrac{k_{12}k_{23}}{k_{12}+k_{21}},\\
\beta\simeq 1+\dfrac{k_{12}k_{23}}{k_{31}(k_{12}+k_{21})}.
\end{gather}
The fit to the measured $g^{(2)}$ functions determine the physical parameters $k_{ij}$ that are summarized in Table~\ref{kij}. Instead of showing $k_{ij}$, we show the corresponding lifetimes $\tau_{ij}=1/k_{ij}$. The first line in Table~\ref{kij} comes from the $g^{(2)}$ measurement when the tip is above the glass. When the tip is facing silver (second line of Table~\ref{kij}), we calculate the mean value for each parameter in all the measurements described before (i.e. Figs.~\ref{G2}(b-d)). This choice comes from the fact that all the measurements on the metal film show nearly the same $g^{(2)}$ antibunching profile (see Fig.~\ref{G2}), and so the parameters that come out are very close.  In figure \ref{G2}, we add the theoretical $g^{(2)}$ function obtained with these mean values and Eqs.~3-6, showing a good agreement between data and the model.  The main fact to stress is that $\tau_{21}$ is six times less when facing silver than in front of glass. We could at first sight assume that it comes from an increase in the excitation power, but looking at the  quantum yield $Q=k_{21}/(k_{21}+k_{23})$ we see that  it goes from 27\% on glass to 74\% (mean value) on silver (see Table~\ref{kij}). Therefore, it is reasonable to assume that the NV center emission is enhanced by the presence of silver. It is important to emphasize that by working in a high excitation rate regime we obtain $Q=27\% \ll 1$ on glass  ~\cite{Martin2}. It is only in such a regime that the SPP coupling can become a benefit when going to the silver side since it can increase $Q$ and reduce significantly the spontaneous emission life-time.\\ 
\begin{table}[htbp]
\centering
\caption{\bf Photophysical parameters. The first line corresponds to the $g^{(2)}$ measurement when facing glass. The second line is the mean value of each parameter for every measurement done on silver.}
\begin{tabular}{cccccc}
\hline
Configuration & $\tau_{21}$(ns)  & $\tau_{12}$(ns) & $\tau_{23}$(ns)& $\tau_{31}$(ns) & $Q$(\%) \\
\hline
 Facing ~glass  & $60$ &$51$ & $23$ &$300$ & $27$\\
 Facing ~silver  & $9.7$ &$27$ & $27.4$ &$102$ & $74$ \\
\hline
\end{tabular}
  \label{kij}
\end{table}
\indent Going back to  the physical interpretation and implication of $g^{(2)}(\tau,\mathbf{k}_A,\mathbf{k}_B)$ measurements,  two points must be underlined.  First, since the correlation curves $g^{(2)}(\tau,\mathbf{k}_A,\mathbf{k}_B)$, $g^{(2)}(\tau,\mathbf{k}_A,\mathbf{k}_A)$ and $g^{(2)}(\tau,\mathbf{k}_B,\mathbf{k}_B)$ are experimentally identical, the isotropy of the photon/plasmon emission by the NVs is confirmed. This is expected since the number of emitters $N=10$ washes out any anisotropy in the probability emission and correlation in the momentum $\mathbf{k}$-space. This is not necessarily the case if we can tailor specific plasmonic antennas and devices selecting particular directions of propagation for SPPs. The methodology proposed here offers therefore huge potentialities for future development and interesting quantum optics experiments involving spatio-temporal second-order coherence. Second, from  a fundamental point of view, the present experiment can be seen as a quantitative wave-particle test involving quantized SPPs. Since $g^{(2)}(0,\mathbf{k}_A,\mathbf{k}_B)<1$ the correlations observed in the Fourier  space cannot be explained using a pure wave-like approach (which is nevertheless necessary to explain the point spread function interferences  shown in Fig.~\ref{CCD}(a) and the plasmonic ring observed in Fig.~\ref{CCD}(b)). We need clearly a dual picture to understand the cross correlations observed in the Fourier plane and the wave-like features of Fig.~\ref{CCD}. Furthermore, we here clearly work with genuine canonical conjugate observables such as momentum $\mathbf{k}$ and position $\mathbf{x}$ for SPPs. Therefore, we are very close to textbook discussions developed to illustrate cornerstone Bohr's principle of complementarity which states that one cannot in a single experiment record for the same particles the statistics associated with incompatible, i.e, non-commutable quantum observables. We remark that whereas in usual examples interference fringes are tested in the Fourier space and path information in the direct space, we have here the opposite situation: the complementarity between SPP fringes is probed in the direct plane whereas the `which-way' information is probed in the Fourier plane.\\
\indent To conclude, we have presented an experimental methodology to study and analyze spatio-temporal second-order correlation for SPPs that is based on LRM and NSOM involving  NV quantum emitters.  We have explained how to observe quantum correlation in both the Fourier and direct planes. This opens the way to for genuine tests of wave particle duality at the single SPP level based on conjugate canonical variables $\mathbf{k}$ and $\mathbf{x}$. The methodology was here illustrated with simple examples involving a uniform thin metal film and an emsemble $N=10$ quantum emitters. However, the approach is clearly not limited to these proofs of principle and offers very interesting potentialities for new experiments involving spatio-temporal second-order correlations for SPPs.  We expect therefore that this work will open up future studies at the interface between quantum plasmonics and near-field optical microscopy.                                

\section*{Funding Information}
\textbf{Funding.} This work was supported by Agence Nationale de la Recherche (ANR), France,
through the SINPHONIE and PLACORE grants.
\section*{Acknowledgments}
\textbf{Acknowledgment.} We thank  J.-F.~Motte, from
NANOFAB facility at Institut Neel for the optical tip manufacturing. We thank T.~Gacoin, from PMC laboratory at Ecole Polytechnique, and G.~Dantelle, from Institut Neel, for providing us the nanodiamonds used in this work.

\end{document}